\documentstyle[10pt,aaspp4,flushrt]{article}
\def\etal{\it et al. \rm }
\begin{document}
 
\title{Cluster Populations in A115 and A2283}

\author{Karl D. Rakos}
\affil{Institute for Astronomy, University of Vienna, A-1180, Wien, Austria;
rakosch@astro1.ast.univie.ac.at}

\author{James M. Schombert}
\affil{Department of Physics, University of Oregon, Eugene, OR 97403;
js@abyss.uoregon.edu}

\author{Andrew P. Odell}
\affil{Department of Physics and Astronomy, Northern Arizona University, Box 6010,
Flagstaff, AZ 86011; andy.odell@nau.edu}

\author{Susanna Steindling}
\affil{Wise Observatory and the School of Physics and Astronomy,
Tel Aviv University, Tel Aviv, Israel; susan@wise1.tau.ac.il}

\begin{abstract} 

This paper presents four color narrow-band photometry of clusters A115
($z=0.191$) and A2283 ($z=0.182$) in order to follow the star formation
history of various galaxy types.  Although located at similar
redshifts, the two clusters display very different fractions of blue
galaxies (i.e. the Butcher-Oemler effect, $f_B = 0.13$ for A115, $f_B =
0.30$ for A2283).  A system of photometric classification is applied to
the cluster members that divides the cluster population into four classes
based on their recent levels of star formation.  It is shown that the blue
population of each cluster is primarily composed of normal starforming
(SFR $< 1 M_{\sun}$ yrs$^{-1}$) galaxies at the high luminosity end, but
with an increasing contribution from a dwarf starburst population below
$M_{5500}= -20$.  This dwarf starburst population appears to be the same
population of low mass galaxies identified in recent HST imaging (Koo
\etal 1997), possible progenitors to present-day cluster dwarf
ellipticals, irregulars and BCD's.  Deviations in the color-magnitude
relationship for the red galaxies in each cluster suggest that a population
of blue S0's is evolving into present-day S0 colors at this epoch.  The
radial distribution of the blue population supports the prediction of
galaxy harassment mechanisms for tidally induced star formation operating
on an infalling set of gas-rich galaxies.

\end{abstract}

\section{INTRODUCTION}

The investigation of the blue and red populations in rich clusters have
been our most lucrative glimpse into the evolution of galaxies.  The
search for changes in the stellar population of galaxies has mostly
focused on clusters of galaxies due to their high visibility and ease of
cataloging rich clusters even at distant redshifts.  However, the
gravitational clumping of a cluster also minimizes the effort required for
distance determination where the measurement of a few of the brightest
galaxies provides the redshift for the entire cluster.  Attention
concerning recent and rapid evolution has been diverted in the past decade
to field galaxies (Tyson 1988), and the dilemma of the blue field
population.  However, clusters are the sites of numerous dramatic
evolutionary effects, such as galaxy cannibalism (Moore \etal 1996) and
the Butcher-Oemler population (Oemler, Dressler and Butcher 1997).  In
addition, cluster populations are key to understanding galaxy
characteristics since they cover not only a full range of galaxy masses
(from giants to dwarfs) but also the full range of Hubble types and
intrinsic density (e.g. surface brightness) that are missing from field
populations.

In a series of papers extending over the last 12 years (Rakos \etal 1988,
1991, 1995, 1996, 1997, 1999), we have used a narrow band filter system to
perform photometry of galaxies in rich clusters for redshifts ranging from
0.2 to 1.  Our studies have differed from previous photometry of distant
clusters by the use of a color system surrounding the 4000\AA\ break (the
Str\"omgren $uvby$ system) and modified such that the filters are
``redshifted'' to the cluster of galaxies in consideration.  This method
results in effectively no k-corrections and allows discrimination between
cluster membership based on spectrophotometric criteria.  We call our
modified system $uz,vz,bz,yz$ to distinguish it from the original $uvby$
Str\"omgren system and we believe we have demonstrated that these color
indices are a profitable tool for investigating color evolution of both
the red and blue populations in clusters of galaxies.

In an previous exploratory paper (Rakos, Maindl and Schombert 1996,
hereafter RMS96), the modified $uz,vz,bz,yz$ Str\"omgren system was used
to develop a photometric classification scheme based on the recent star
formation history of a galaxy.  From the $mz$ color index
($mz=(vz-bz)-(bz-yz)$), the classification technique was shown to be
extremely successful at discriminating normal starforming galaxies
(spirals) and starburst galaxies despite the presence of heavy reddening.
Rakos \etal (1996, 1997) applied this technique to the members of the
blue population in several clusters of intermediate redshift ($0.2 < z <
0.6$) and demonstrated that many of the cluster members have strong
signatures of star formation activity.  An extreme example is CL0317+1521
at $z=0.583$ which has a blue fraction of $f_B=0.60$ and which 42\% of the
blue population have $mz<-0.2$, the photometric signature for a
starburst.  Deep photometry of the cluster A2317 ($z=0.211$, Rakos, Odell
and Schombert 1997) demonstrated that the ratio of the blue population to
red population has a strong dependence on luminosity, such that blue
galaxies dominate the very brightest and very faintest galaxies in the
cluster.  In contrast to the bright blue galaxies, the fraction of
galaxies displaying the signatures of a starburst increases towards the
faint end of the luminosity function, a dwarf starburst population first
suggested by Koo \etal (1997).

The origin of this dwarf starburst population remains an enigma.  Tidal
interactions are frequently invoked as an explanation for the high
fraction of starburst galaxies in Butcher-Oemler clusters (Dressler \etal
1994, Couch \etal 1994).  If the same phenomenon acts on the low mass
galaxies, then these starburst systems would have their origin as gas-rich
dwarf galaxies who have had a short, but intense, tidally induced episode
of star formation which would quickly exhaust their limited gas supply.
It should be noted, however, that the orbits of cluster galaxies are
primarily radial (Bothun and Schombert 1990), and the typical velocities
into the dense cluster core are high.  This makes any encounter extremely
short-lived, with little impulse being transfered as is required to shock
the incumbent molecular clouds into a nuclear starburst.  The galaxy
harassment mechanism (Moore \etal 1996) emphasizes the influence of the
cluster tidal field and the more powerful impulse encounters with
individual central galaxies at the cluster edges.  These two
processes can then conspire raise the luminosity of cluster dwarfs,
increase their visibility and, thus, their detectability.

To further explore the behavior of the red E/S0's, blue Butcher-Oemler
galaxies and the newly detected dwarf starburst population, we began a
program of obtaining deeper photometry then our previous work on
intermediate redshift clusters.  Our goal in this study is to 1)
illuminate the types of galaxies involved in the Butcher-Oemler effect, 2)
determine the dominance of the blue galaxies to the cluster luminosity
function, 3) examine the spatial extent of the blue and red population,
and 4) determine the existence and characteristics of the dwarf starburst
population in other blue clusters.  In order to maintain a correct
timescale to stellar population models, calibrated to globular cluster
ages, values of $H_o=50$ km sec$^{-1}$ Mpc$^{-1}$ and $q_o=0$ are used
throughout this paper.

\section{OBSERVATIONS}

\subsection{\it Cluster Sample}

To determine if the results for A2317 were accidental, we have expanded
our program to three clusters at similar redshifts; A115 ($z=0.191$),
A2283 ($z=0.182$) and A2218 ($z=0.180$).  Table 1 shows the cluster
coordinates (1950.0), the Bautz-Morgan classification and the background
corrected count $C$ in the magnitude range $m_3+3$ (Abell richness).  A115
and A2283 are similar to A2317 in richness.  A2218 is, on the other hand,
one of the richest clusters in the Abell catalog and contains a large cD
galaxy in its center which extends over more than 180 kpc (Wang and Ulmer
1997).  A115 is also a known binary cluster (Beers, Huchra and Geller
1983) which is normally associated with a dynamically young system.  A115
was also studied by our group in a previous paper (Rakos, Fiala and
Schombert 1988); however, only 53 galaxies were measured at that time.
This paper presents new results for A115 and A2283.  The data for A2218
will be presented in a later paper (Rakos, Dominis and Steindling 2000).

The observing procedure and reduction used herein is similar to that
described in Rakos and Schombert (1995).  The observations of A115 were
taken with 90-inch Steward Observatory telescope using 800x1200 pixel CCD
and a focal reducer with 600 arcsec circular unvignetted field at a f/9
input beam.  A2283 was observed with the 4m KPNO PFCCD Te1K, the 90-inch
Steward telescope in direct mode plus 2K CCD and the 1m Wise telescope in
Israel using 1K TEK CCD.  The field size for the Wise telescope is 10
arcmin square.  Calibration to the $uz,vz,bz,yz$ system used
spectrophotometric standards (Massey \etal 1988).  Colors and magnitudes
were measured using IRAF APPHOT and are based on metric apertures set at
32 kpc (9 arcsecs for A115 and A2283) for cosmological parameters of
$H_o=50$ km sec$^{-1}$ Mpc$^{-1}$ and $q_o=0$.  Our typical internal
errors were 0.05 mag in colors at the bright end of the sample and 0.1 mag
for the faint end.  No corrections for Galactic extinction were made due
to the high latitude of the clusters.  All the photometry ($uz-vz$,
$bz-yz$, $vz-yz$ and $mz$) is available at the narrow band photometry web
site (zebu.uoregon.edu/$\sim$js/narrow).

Photometry using 32 kpc apertures was performed on the final, co-added
frames.  Objects were selected based on detection in all four filters at
the 3$\sigma$ level.  In addition to the colors, the $yz$ filter values
can be converted to $m_{5500}$.  The Str\"omgren system was originally
designed only as a color system; however, the $yz$ filter is centered on
5500\AA\ and it is possible to link the $yz$ flux to a photon magnitude,
such as the AB79 system of Oke and Gunn 1983, through the use of
spectrophotometric standards.  This procedure was described in detail in
Rakos, Fiala and Schombert (1988) and is used to determine the $m_{5500}$
values for all the cluster members.  Typical errors were 0.02 mag for the
brightest cluster members to 0.08 mag for objects below $m_{5500}=20$.
Incompleteness of both clusters is evident below $M_{5500}=-20$, the
faintest objects are $M_{5500}=-19$.

\subsection{\it Cluster Membership}

Each of the filters used for this study are approximately 200\AA\ wide and
were specially designed so that their central wavelengths matches the rest
frame of the cluster.  This provides a photometric method of determining
cluster membership, without the use of redshift information, due to the
unique shape of any galaxy's spectra around the 4000\AA\ break.  Our
method is fully discussed in Rakos, Schombert and Kreidl (1991) and Fiala,
Rakos and Stockton (1986), but a brief discussion follows.

Figure 1 displays the changes in color indices for a standard elliptical
profile (NGC 3379, Kennicutt 1992) and a starforming disk (NGC 6643,
Kennicutt 1992) as a function of redshift.  Since the 4000\AA\ break
brackets the $uz$ and $vz$ filters, sharp blue $vz-yz$ colors are seen for
low redshift objects (relative to the cluster redshift, i.e. foreground)
and red $bz-yz$ colors are found for high redshift objects (background).

Our procedure to determine cluster membership is to use the color
information provided by the $mz$, $vz-yz$ and $bz-yz$ indices to measure a
distance from the normal galaxy templates (see Rakos, Schombert and Maindl
1996).  Any object more than 3$\sigma$ from the mean relationships is
eliminated from the sample.  As expected, a majority of the rejected
objects are too red (background), but the criteria also eliminates
foreground stars.  While there could be contamination from poor S/N
galaxies, whose erroneous colors place them within the 3$\sigma$ boundaries.
However, these objects are eliminated by our photometric quality criteria
(typically $\sigma < 0.07$ mag, see below).

To test this procedure for A115, redshifts were extracted from NED.
Within the 10 arcmin field of A115, 14 galaxies with redshifts are found,
12 at the cluster redshift ($z=0.191$) and two at a foreground redshift of
14,388 and 14,307 km sec$^{-1}$.  Ten of the twelve cluster galaxies were
correctly identified by the photometric criteria, two others were
eliminated due to poor S/N.  The two foreground galaxies were correctly
eliminated as non-cluster members by their $mz$ indices.

In A2218, five galaxies have redshifts, two cluster members and three
foreground objects.  As with A115, all three foreground objects were
eliminated by the photometric criteria.  One of the cluster members was
identified, the other was too close to a bright, nearby galaxy for proper
measurement.  Combined with the A115 results, we believe this to validates
our procedure, at least for the brighter cluster members.  We only have
numerical simulations for the fainter cluster members.

\subsection{\it Photometric Classification}

Before the advent of 2D photometry and spectroscopy, galaxies were
understood by their morphological appearance and integrated colors.  Star
formation, in particular, was estimated based on the size and number of
H\,II regions rather than the number of ionizing photons as produced by a
set number of O and B stars.  Currently, the morphological appearance of a
galaxy has become a secondary parameter as the field of galaxy evolution
has become more concerned with the actual star formation history of
galaxies based on fits to spectrophotometric models, rather than the
inferred history based on our conception of how star formation and galaxy
appearance are related.  Although there is a strong correlation between
the current star formation rate and the morphological class of a galaxy,
most of the questions arising in high redshift cluster studies will
involve the underlying stellar and gas components of a galaxy, not its
appearance.

Classification by colors or spectral type have their origin in the
Morgan system (Morgan and Mayall 1957).  Although primitive in its
information, the Morgan system was key to developing an understanding of
the morphological type of a galaxy as it relates to the state of its
underlying stellar population.  Modern studies have successful applied
spectrophotometric classification to starforming and starburst systems in
distant clusters (Poggianti \etal 1999).

The exact classification procedure for our Str\"omgren filters was
discussed in RMS96, but the following is a brief review.  For the purposes
of classification, the galaxy types in clusters are basically divided into
four categories based on their color indices as they reflect into current
and recent star formation.  The four classes are 1) a passive,
non-starforming object (presumably the counterpart to present-day
ellipticals and S0's, at least in terms of the current star formation rate
(SFR), if not the integrated past SFR), 2) a normal starforming object
(with an SFR between 0.1 and 1 $M_{\sun}$ yrs$^{-1}$ and standard values
for the IMF, Kennicutt 1998) which can be associated with present-day
spirals, 3) an object with anomalous colors that correspond to a region in
the two-color plane occupied by starburst objects (i.e. SFRs greater than
1 $M_{\sun}$ yrs$^{-1}$) and 4) objects with a non-thermal component to
their luminosity that is associated with Seyfert activity.

This classification system is not exact, there are several transition
regions that produce an ambiguous assignment of the state of the
stellar population.  The three most serious are the regions occupied
by early-type spirals, low luminosity Seyfert galaxies and weak or
post-starburst systems.  Early-type spirals are found to overlap with
the passive or E/S0 region of photometric space.  This is not surprising
since our system emphasizes the integrated color of a galaxy, and the
large bulge component of early-type galaxies dominates the colors over a
minor disk component.  The impact of early-type spirals is
minimized within the goals of our work since we can easily group galaxies
with short gas depletion rates (ellipticals, S0's and early-type spirals)
into the same class when the focus is star formation rates.

The Seyfert classification overlaps with normal spirals depending on the
ratio of the amount of light from the AGN versus the galaxy stellar disk.
Most Seyfert galaxies are spiral (Ho, Filippenko and Sargent 1997), and
heavily weighted towards late-types, in agreement with the model that AGN
luminosity is correlated with the ability to fuel the central engine with
HI gas.  Since a global color measurement will overemphasize the
contribution from normal starlight with respect to non-thermal emission,
AGN colors will typically scatter towards the normal spiral sequence in
multi-color space.  Thus, when interpreting the fraction of Seyferts in
our cluster sample, we must place the caveat that this fraction represents
the number of `extreme' Seyferts, i.e. ones where the light from the AGN
is much stronger then the total disk and bulge contribution.

Lastly, the classification of galaxy as a starburst is based on a
comparison with an IRAS sample of interacting galaxies and starburst
models by Lehnert and Heckman (1996).  Even highly reddened starburst
systems are easily distinguished for ellipticals in multi-color diagrams
such as $vz-yz$ versus $bz-yz$ (see RMS96).  However, the tail end of the
spiral sequence, composed of the late-type spirals, overlaps the starburst
region.  Again, from the point of view of a classification system aimed at
the underlying stellar population, this simply acknowledges the fact that
many late-type spirals have SFRs high enough to warrant a starburst
classification.  Thus, we emphasize that our scheme is not morphological
and a majority of galaxies with Sc or Sd appearances would have a
photometric classification of starburst in our scheme.

To increase the reliability of photometric classification, particularly
for those galaxies whose colors place them at the borders of two different
classifications, we have explored using data from nearby galaxies as
template colors and assigning classification based on a statistical match
in all four colors ($mz$, $uz-vz$, $vz-yz$, $bz-yz$).  To this end, 132
spectrophotometric scans were extracted from the literature as outlined in
Rakos, Maindl and Schombert 1996.  The templates are selected from four
catalogs: 1) elliptical/S0 (Gunn and Oke 1975), 2) normal spirals
(Kennicutt 1992), 3) Markarian (De Bruyn and Sargent 1978) and an 4) IRAS
sample (Ashby, Houck and Hacking 1992).  All four of these catalog samples
are shown in Figure 2.  The division for the IRAS sample is particularly
striking considering the sample was selected solely for far-IR color.
IRAS galaxies with merger signatures (disturbed morphology, tidal
features) are indicated by the solid symbols, but do not distinguish
themselves from the other IRAS galaxies, probably an indication that
dynamical features have much shorter timescales than star formation
effects (Sanders \etal 1988).  The Markarian sample is a good example of
the degree of difficulty in classification when there is a mixture of
non-thermal and disk colors as discussed above.  In this case, there is a
clear extension of the Markarian colors to anomalous $bz-yz$ values for
constant $mz$, but the overlap with normal spirals is problematic.  In the
overlap region, classification as a spiral is preferred in order to trace
the global star formation history of the galaxy, rather than the evolution
of the nuclear region.  The spiral sequence neatly divides into early and
late-types, with early-types encroaching on the elliptical region and
extreme late-types displaying starburst colors, again, reflecting the
emphasis on star formation effects rather than dynamical or morphological
appearance.

The multi-color, rest frame nature of our datasets have allowed for a
fairly accurate classification of distant cluster members by photometric
means since no k-corrections are involved.  No single multi-color diagram
(such as Figure 2) can portray all the information contained in the full
photometry for each galaxy.  Particularly for galaxies in the border
regions, small errors in a single color can produce an erroneous
classification.  In order to minimize these errors, photometric class is
assigned by a statistical weighting with respect to the template data.
The templates are the same galaxies used in RMS96, only now weighted for
all their colors (see Steindling, Borsch and Rakos 2000 for further
discussion).  The distant cluster data is directly compared with the
template data and classification used herein is assigned based on the
closest match.  The difference between this method and simply using the
divisions sketched RMS96 are minor, but this statistical method allows for
some estimate of the internal errors of classification.

\subsection{\it A115}

A115 is a binary cluster (Beers, Huchra and Geller 1983), our CCD survey
centered on the southern component.  The field of view does not include
the northern component, so the data should be considered a study of only
one subcluster in A115.  Over 300 objects in the A115 fields were detected
in all the four filters.  For final analysis, a photometric selection
criterion of cluster membership was applied such that an object must have
an internal accuracy of $\sigma < 0.07$ mag to be included.  This resulted
in a total of 100 cluster members.  The classification fractions for both
the blue and red populations are listed in Table 2.  A115 was also studied
on one of the original papers of this series (Rakos, Fiala and Schombert
1988).  The current sample triples the number of galaxies with narrow band
photometry for A115 from our original work, and the accuracy is improved
such that we can now sample below $L^*$.  The external errors were less
than 0.01 for the lowest luminosity galaxies, and the mean color of the
cluster is now $vz-yz=0.527$ compared to the old value of 0.529.

Figure 3 presents the $vz-yz$ versus $bz-yz$ and $mz$ diagrams for the 100
cluster members.  The various symbols display the photometric
classifications based on template comparisons.  The solid line in the left
panel Figure 3 displays the relationship for normal spirals (Rakos, Maindl
and Schombert 1997).  Ellipticals and S0's have a slightly shallower
correlation due to the color-magnitude relation (see \S3.1).  The
starburst objects near the $mz$ cutoff (blueward of $vz-yz=0.1$) probably
represent starforming late-type spirals.  The red starburst systems lie
along the reddening vector, below the spiral sequence.  Most of the
starburst systems with $mz < -0.25$ are low in luminosity (and, therefore,
mass), and are the reddest of the non-elliptical objects in the sample.
This starburst population is similar to the dwarf starburst population in
A2317 (Rakos, Odell and Schombert 1997).  The solid lines in the right
panel of Figure 3 display the mean cutoffs for the population types,
although comparison to templates is used for the final classifications as
described in \S2.2.  

The population fractions are fairly normal for an intermediate redshift
cluster.  We find that 68 (68\%) of the cluster members are type E/S0,
whereas 11 (11\%) are Sp/Irr, 17 (17\%) are starburst and remaining 4
(4\%) are classified as Seyfert galaxies.  These ratios are slightly
richer in early-type galaxies compared to the ratio for a nearby rich
clusters, given as E:S0:Sp/Irr ratios of 20\%:40\%:40\% (Oemler 1992).  It
is assumed that all the galaxies photometrically classified as Sp/Irr,
starbursts and Seyferts would be morphologically classified as late-type
category.  From this point of view, A115 is more evolutionary
``developed'' than A2317, meaning that it has a richer, red galaxy
population and a weaker current star formation rate for the cluster as a
whole.  This is surprising given the dynamically young appearance to the
A115 system as a whole (Beers, Huchra and Geller 1983).  The fraction of
Seyferts found in local clusters is extremely low, less than 1\% of the
total cluster population.  However, a recent study by Sarajedini \etal
(1996) finds the fraction of AGNs to increase to 10\% between $z=$0.2 and
0.6.  Thus, our observed value of 4\% is in good agreement with expected
number of AGNs at this epoch.

The blue fraction ($f_B$) for A115 is 0.13 (again, we note that the values
presented herein are for the southern subcluster only), where we maintain
the original definition of blue populations from Butcher and Oemler (1984)
as the number of galaxies 0.2 mag blueward from the k-corrected mean E/S0
color.  In the Str\"omgren system, 0.2 mag from the E/SO line translates
into a rest frame blue/red cutoff of $bz-yz=0.2$.  This criteria will
place many of the early-type spirals in the red category, although this
would be expected from the dominance of the red bulge in such galaxies.
However, a significant number of the starburst class galaxies will be
misidentified as red population members due to heavy reddening by dust.
As we will see below, many of the red starburst objects are low in
luminosity and would not have contaminated the blue fraction values from
earlier work of Butcher and Oemler (1984) or Dressler and Gunn (1983).

\subsection{\it A2283}

About 350 objects were measured in the 10 arcmin square field surrounding
A2283.  Elimination of foreground and background objects, stars and
applying the photometric selection criterion for a minimal error of 0.07
mag, gives a total of 79 cluster members.  The population fractions are
listed in Table 3, such that 42 (53\%) of the cluster members are E/S0,
whereas 16 (20\%) are Sp/Irr, 15 (19\%) are classed as starburst and
remaining 6 (8\%) are classified as Seyfert or AGN galaxies.

Figure 4 presents the $vz-yz$ versus $bz-yz$ and $mz$ diagrams for the 79
cluster members.   Symbols and lines are the same as Figure 3.  As can be
seen in the figure, A2283 has a broader population of spirals as compared
to A115.  There is slightly more scatter about the
spiral relation, having colors blueward about 0.1 mag for their $vz-yz$
colors.  This is probably due to weak AGN colors contaminating any
otherwise pure disk population.  The bluest starburst systems are redward
of the spiral sequence along the reddening vector.  The slope of the
elliptical/S0 population in the $vz-yz$, $bz-yz$ plane is the same as
A115.

The population of A2283 is richer in Sp/Irr and starburst galaxies
compared to A115.  We find that 39 (49\%) of the cluster members are type
E/S0, whereas 21 (27\%) are Sp/Irr, 17 (22\%) are starburst and remaining
2 (2\%) are classified as Seyfert galaxies.  This is slightly more
spiral-rich than the nearby cluster morphological ratios, assuming that
the starburst systems would be classed as late-type galaxies.  The
fraction of blue to red galaxies is 30\% ($f_B=0.30$), high compared to
A115 but similar to A2317 (Rakos, Odell and Schombert 1997).  For this
reason, we refer to A115 (and A2218) as red clusters and A2283 and A2317
as blue clusters.  Again, the blue galaxies dominate the very bright and
very faint end of the luminosity function and are located primarily at the
cluster edges, which also mimics the behavior of A2317.

\section{DISCUSSION}

\subsection{\it Cluster Population Fractions}

The results of the photometric classification for A115 and A2283 are shown
in Figures 3 and 4 and summarized in Tables 2 and 3.  The distribution of
the three primary galaxy types (ellipticals/S0s, spiral/irregular and
starburst), as a function of magnitude, are also shown in Figure 5 for the
combined populations of A115, A2283 and A2317.  Although the three
clusters differ in the fraction of their blue, red and starburst
populations, all three follow the same behavior with respect to population
type and luminosity.  The ellipticals/S0 population have numerical
superiority in all three clusters and have a distribution of luminosity
that is typical for rich cluster of galaxies (Dressler \etal 1999).  In
contrast, the photometrically classified spirals/irregulars have a sharply
different distribution, peaking at high luminosities ($M_{5500} \approx
-21$) and falling off to fainter magnitudes.  The galaxies classified by
our photometric scheme as starburst display the opposite behavior, rising
to become a majority of the cluster population below $M_{5500}=-21$.  

Where the behavior for the different photometric types in A115, A2283 and
A2317 seems contrary to our expectations from studies of nearby clusters
(i.e. that ellipticals and S0's dominant the bright end of luminosity
function), it is not surprising when we remember that this is a
photometric classification scheme.  The objects in the spiral/irregular
category may not be morphologically disk galaxies.  Their classification
is only a measure of the recent star formation history.  In fact, the same
luminosity distributions are evident in the Dressler \etal (1999)
spectrophotometric catalog of ten distant clusters.  Figure 7 of that
paper displays the luminosity distribution for their sample divided by
physical morphology and spectrophotometric class.  The morphological
distributions display the familiar cluster phenomenon of the bright end of
the luminosity function being rich in ellipticals and S0's and late-type
galaxies having lower mean luminosities.  The number of spirals and
irregulars is higher in the Dressler \etal sample compared to nearby
clusters, and there are a larger number of bright, blue late-type galaxies
(the Butcher-Oemler effect).  The early-type galaxies still comprise a
majority of the brightest galaxies.  However, when the sample is divided
by spectrophotometric class (where their k/k+a corresponds to our E/S0,
e(a)/e(c) corresponds to our S/Irr and e(b) corresponds to our starburst
class) then the luminosity distributions agree remarkably well with our
Figure 5.  Their e(a)/e(c) classes peak at $M_{5500} = -23$ ($H_o=50$ km
sec$^{1}$ Mpc$^{-1}$) and decrease in number, the same as our
spiral/irregular class.  Their e(b) class also follows the same behavior
as our starburst population, a slow rise starting at $M_{5500} = -22$.

The differences between the characteristics of the three photometric
populations can also be seen in Figure 6 where the fractions of the
ellipticals/S0s, spiral/irregulars and starbursts galaxies are divided
into four luminosity bins.  The top panel displays the data for A2283 and
A2317, the two clusters with high $f_B$ values.  The bottom panel displays
the data for the red cluster, A115.  Immediately obvious from Figure 6 is
the fact that there are two competing contributions to the number of blue
galaxies in all three clusters.  At the bright end, one finds luminous
objects with normal star formation rates which we classify photometrically as
spirals/irregulars.  As the contribution by this spirals/irregular
population declines with luminosity, the fraction of starburst systems
increases.  This trend is common to both the red cluster (A115) and the
blue clusters (A2283 and A2317).  It seems clear from this figure that the
Butcher-Oemler effect, the increase in $f_B$ at these redshifts, is
primarily due to the contribution of galaxies with spiral-like colors
(i.e. normal star formation rates).

For further analysis of the changes in galaxy photometric type, we have
divided the cluster members into three sub-populations; red, blue and
starburst.  The division between blue and red uses the same Butcher-Oemler
criteria ($bz-yz<0.2$) as was used to determine $f_B$ in each cluster.
All objects classified as starburst, based on their $mz$ index, comprise
the starburst population.  Over 98\% of the galaxies classed as
elliptical/S0 (non-starforming) are members of the red population.  The
galaxies classed as Sp/Irr are evenly split between the red and blue
populations (probably due to the effect that the early-type spirals by
morphology comprise most of the red population Sp/Irr's).  The starburst
galaxies would also be fairly evenly split into blue and red types, if not
separated into their own class.  Each of these three sub-populations will
be examined in the next sections.

\subsection{\it Red Population}

One of the dilemmas associated with the discovery of the Butcher-Oemler
effect is that a majority of the present-day cluster population is
composed of old, red objects (ellipticals and S0s) with effectively no
recent SFR.  However, a large number of the cluster galaxies at
intermediate redshifts are involved in star formation at relatively high
levels.  The mystery, then, revolves around whether the blue population
individuals are converted into red members with time or whether blue
cluster members fade from view with time (or destroyed), boasting the
ratio of red galaxies.  All indications, from studies of the age and star
formation history of present-day cluster ellipticals (Abraham \etal 1999,
Buzzoni 1995) and the color evolution of the red population
to redshifts of 1 (Rakos and Schombert 1995), are that the history of
ellipticals is one of passive evolution from an epoch of formation near a
redshift of 5.  Where there may be transitions at intermediate
redshifts of blue S0's into red S0's (as gas is depleted from star
formation, see discussion below), a majority of the red population must
have its origin from galaxies who have undergone an initial burst of star
formation followed by passive evolution and, thus, should not be a member
of the blue population at redshifts of 0.4.

This being the case, the red population provides an opportunity to examine
age and metallicity predictions as a function of time.  The most obvious
test is to explore the mass-metallicity relationship of the red population
as reflected in the observed color-magnitude diagram.  In the closed-box
models of galaxy evolution (Faber 1973; Tinsley 1980), the greater the
mass of a galaxy (higher luminosity), the higher its supernova rate which
results in a more rapid initial enrichment of the first generation of
stars.  This, in turn, produces a redder red giant branch (RGB) for the
composite population and strong signatures in the $vz-yz$ indices.
The $vz$ filter is located directly over several strong metal lines in an
old population spectrum (Ca H and K, G band and Mg).  Thus, there is the
expectation that the $vz-yz$ colors should be more sensitive to
metallicity changes than most broadband indices.  According to Visvanathan
and Sandage (1977), the color magnitude relation is nearly zero from 5000
to 7000\AA\ and increases steadily shortward of 4300\AA\ reaching a
maximum value of 0.11 mag per mag near 3400\AA.  The characteristics of
metallicity and the Str\"omgren colors used here was studied extensively
in Schombert \etal (1993) which found, and was confirmed by
spectrophotometric models, that the $vz-yz$ color was the strongest metal
indicator for old stellar populations (i.e. ellipticals and S0's) of the
$uz,vz,bz,yz$ indices, and will be used for our color-magnitude analysis.

The $vz-yz$ color-magnitude relation is shown in Figure 7 for the red
populations in A115 (68 objects) and A2283 (39 objects) along with a
linear fit.  For this test, only the members of the red population
classified as ellipticals/SOs are used and, despite the radically
different population fractions between the two clusters, there is no
statistically significant difference in the color-magnitude relation
between the two clusters.  Also plotted in Figure 7 is the red population
data for A2317 (Rakos, Odell and Schombert 1997) along with the
color-magnitude relation for present-day ellipticals from Schombert \etal
(1993).  The A2317 data is a excellent match to the present-day
color-magnitude relation ($\Delta(vz-yz)/\Delta(M) = -6.6 \times
10^{-2}$), but the data from A115 and A2283 is a factor of
three shallower in slope ($-9.8 \times 10^{-3} \pm 1.2 \times 10^{-2}$ and
$4.5 \times 10^{-3} \pm 1.6 \times 10^{-2}$ respectfully).  While there is
agreement at the low luminosities, the A115 and A2283 data do not redden
as strongly as the A2317 data for increasing luminosity.  This discrepancy
does not appear to be due to errors in the observations.  The A2317 data
is of similar accuracy as the A115 and A2283 data, plus the shallow slope
does not appear to be due to increased scatter at the low luminosity end.
In fact, the brightest galaxies in both clusters display an unusual amount
of scatter in $vz-yz$ for their luminosity, despite the fact that their
errors are the smallest.  Thus, we conclude that the differences in the
red populations are intrinsic to the clusters.

Closer inspection of the color-magnitude diagrams for A115 and A2283 shows
that A2283 has a similar distribution as A2317, but with a larger scatter
that lowers the slope as calculated from a regression fit.  A115, on the
other hand, clearly has a large number of bright galaxies, classified
photometrically as ellipticals/S0s, with colors that are too blue for
their luminosity, lying both below the red population in A2317 and below
the relationship derived from present-day ellipticals.  It should be
noted that A115 has several galaxies that are in excess of the brightest
cluster members of either A2283 or A2317 which contribute heavily to this
effect.

While it would be interesting to speculate that the shallower slope in the
color-magnitude relation is due to some evolutionary effect, comparison to
models (Buzzoni 1995) indicates that subtle changes in the underlying
stellar population dominate over metallicity effects at this epoch.  For
example, the contribution from horizontal branch (HB) stars peaks at these
redshifts for a galaxy formation epoch between 5 and 10 (see Rakos and
Schombert 1995).  The HB sub-population has a brief dominant
phase in the integrated colors, with an change of 0.1 mag in $vz-yz$
color at redshifts of 0.4 (Charlot and Bruzual 1991), that would serve to
lessen the strength of integrated mass-metallicity effect as only a narrow
slice of stellar mass function is sampled rather than the whole RGB 
population.  And this effect would be amplified if the HB sub-population
has a range of metallicity producing blue to red HB stars, or if mergers
with dwarf, metal-poor galaxies, also containing blue HB stars, has
occurred in the recent past for the brighter cluster galaxies (i.e. galaxy
cannibalism).  This HB effect was seen in the color with redshift plots of
Rakos and Schombert (1995) and would seem to be our best candidate to
explain the larger range in colors for the A115 and A2283.

An alternative explanation is that the greater scatter in $vz-yz$ color is
due to contamination by blue S0's.  These would be S0's which have
recently (in the last 3 to 4 Gyr's) depleted their gas supplies and are
evolving towards elliptical-like colors.  The timescale for such a
transition from spiral to S0 is evident in the near-IR colors of S0 disks
(Bothun and Gregg 1990), and the brightest blue galaxies (classed as
Sp/Irr in our scheme) would be obvious progenitors to these objects.  In
any event, it appears that, in the same manner as the blue population
fraction of the Butcher-Oemler effect is irregular from cluster to
cluster, the color-magnitude relation of the red population vary
dramatically at intermediate redshifts and will require additional
spectrophotometric information before any conclusions can be reached.

\subsection{\it Blue Population}

One advantage of the Str\"omgren colors (or any 4000\AA\ color system) is
the strong dependence of its indices on recent star formation and their
sensitivity to recent deviations in star formation rates that signal starburst
activity.  RMS96 outlines the $uvby$ system sensitivity to star formation
effects using the starburst models of Lehnert and Heckman (1996, see
Figure 6 of RMS96).  Starforming galaxies cover a range of colors from
blue to red with the more extreme starbursts located on the reddened side
of the two color diagram.  For the sake of consistence with previous work,
we have maintained our past definition of $f_B$, the blue fraction in a
cluster, based on a rest frame cut at $bz-yz<0.2$ regardless of extinction
effects.  While this criteria may miss a substantial number of reddened
starburst systems, thus reducing the meaning of $f_B$ with respect to a
clusters star formation history, we can isolate that population with other
indices and discuss them separately (see \S3.4).

The fraction of blue galaxies ($f_B$) in A115 is $0.13\pm0.04$, whereas
the $f_B$ value for A2283 is $0.30\pm0.06$.  This range is not unexpected
as clusters located at intermediate redshifts display a wide variance of
$f_B$ values (from 0.11 in CL0024.5+1653 to 0.65 in A227, Rakos and
Schombert 1995).  In fact, the Butcher-Oemler effect is better described as
the increased occurrence of blue populations with redshift, not a uniform
correlation with redshift (Allington-Smith \etal 1993).  As noted by
the original Butcher-Oemler papers, there exist several distant clusters
with quite red mean cluster colors, meaning similar to present-day
clusters population fractions.

The $f_B$ value in A2283 is similar to A2317 (Rakos, Odell and Schombert
1997) in that the blue population is strong at $f_B=0.30$.  Also similar
to A2317, the blue population in A2283 displays a clear dependence on
absolute magnitude such that blue galaxies dominate the very bright and
very faint end of the luminosity function.  As discussed in \S3.1 and
shown in Figure 6, the blue population at the high luminosity end is
comprised mostly of galaxies photometrically classified as
spirals/irregulars with an increasing contribution by starburst systems at
low luminosities.  Since early detections of the Butcher-Oemler effect
were based on samples heavily weighted towards the bright galaxies, we can
reliably state that the Butcher-Oemler effect, and therefore most of the
blue population, is due to galaxies with normal star formation rates (SFR
$< 1 M_{\sun}$ yrs$^{-1}$).  In clusters with a high $f_B$ values (such as A2283
and A2317), there is a minor blue component from low luminosity
starbursts, but this population has little impact on the calculated values
of $f_B$ since a majority of those systems are heavily reddened.

The blue population in A2317 was found to have a wider distribution
of radii then the red population (see Figure 6 in Rakos, Odell and
Schombert 1997).  Figure 8 displays the same analysis for A2283 and
demonstrates that the same effect is also found there (the blue population
in A115 was too small for a robust statistical test).  The difference in
the radial distribution of the blue population is even more extreme in
A2283 than A2317, having a peak at 0.5 Mpc from the cluster center.  This
result remains regardless if the cluster center is calculated using
geometric cluster center, the x-ray center or the center defined by the
brightest cluster galaxy.  Figure 8 also shows that the red population in
A2283 is strongly peaked at the center with a core density of 160 galaxies
per Mpc$^2$.  The blue population, on the other hand, are deficient in the
cluster core, located primarily in the outer regions of the cluster.

As discussed in Rakos, Odell and Schombert 1997 for the A2317 data, a
mechanism for cluster induced star formation has been proposed by Moore
\etal (1996) called galaxy harassment which emphasizes tidally induced
star formation imposed by both the mean cluster tidal field and
rapid impulse encounters with massive galaxies at the cluster core.  One
of the predictions of galaxy harassment is that galaxies in the cores of
clusters will be older than galaxies at the edges.  In terms of star
formation history, this is the behavior seen in Figure 8 for A2283 and
A2317.  The blue population (i.e. the harassed population) is primarily
located in the outer 2/3's of the cluster.  Other explanations, such as
infall of spiral-rich subclusters, are unsupported since the blue
population is not clumped nor confined to the edges of the cluster.

We arrive at the conclusion that each of the clusters studied herein seem
to be composed of two distinct components in terms of star formation
history.  The scenario for the construction of the blue population
proposed here is similar to the one we proposed in RS95, that the
Butcher-Oemler population is an evolved set of gas-rich (possibly low in
surface brightness) galaxies.  The mechanism of galaxy harassment induces
highly efficient star formation in these dark matter dominated systems
that increases their luminosity and visibility.  Later encounters with the
cluster core destroy these low density systems by tidal forces,
effectively removing them from the present-day cluster sample.
Eventually, the Butcher-Oemler clusters at intermediate redshifts will
evolve into future cD clusters from the released stellar material of the
disrupted blue population.

Lastly, we consider the Seyfert-like galaxies in A115 and A2283.  Although
the number of Seyferts is slightly above normal (as compared to nearby
rich clusters), they are by no means a major component of the
Butcher-Oemler effect.  The amount of AGN activity certainly increases
with redshift for cluster populations (Sarajedini \etal 1996) and the
number of galaxies classified as Seyferts in this data set is in agreement
with those studies even though we believe our classification to be
extremely conservative.  Whether the increase in the number of Seyferts is
related to the increase in the blue population with redshift (the
Dressler-Gunn effect) or whether this is a parallel processes remains
undetermined by this dataset.  A worthwhile future study would be high
resolution spectroscopy of the Seyfert objects and many of the high $mz$
spiral/irregulars as a probe to the range and strength of AGN's in
Butcher-Oemler clusters.

\subsection{\it Starburst Population}

The behavior of starburst galaxies for the $uz,vz,bz,yz$ color system was
investigated in RMS96 by analysis of a set of interacting and merging
galaxies.  In that paper, it was demonstrated that it is possible to
differentiate between starburst galaxies shrouded in dust with strong
reddening (IRAS starbursts, Lehnert and Heckman 1996), versus the strong
ultraviolet radiation of Wolf-Rayet galaxies, simply by the distance the galaxy
lies from the spiral/irregular sequence along the reddening vector,
although the exact amount of reddening is difficult to determine due to
its shallow slope.  The reddening line (see Figures 3 and 4) indicates that
a majority of the starburst galaxies from the IRAS sample have Sb or later
colors with significant extinction ($A_V > 5$) by dust, in same range as
IRAS starbursts (Leech \etal 1989).

Despite the difference in the blue fractions of A115 and A2283, their
starburst populations are similar at 17\% and 22\% respectfully (A2317
also has a 22\% starburst fraction).  Both clusters starburst population
are heavily weighted toward the faint end of the luminosity function, as
was the starburst population in A2317.  The starburst galaxies are evenly
divided between the blue and red populations, reflecting the frequent
occurrence of heavy dust extinction since the blue starbursts lie on the
spiral sequence and the red starbursts lie in the reddened portion of the
two-color diagrams.  For example, most of the blue population in A115 is
due to starburst galaxies since there is a real deficiency in
spiral/irregular class objects.  In contrast, the blue population in A2283
is comprised evenly of both spirals/irregulars and starbursts.

The luminosity distribution indicates that the starburst population
entails a phenomenon that is focused on the faint galaxies in a cluster.
While some of the starburst systems are quite luminous, comparable to Arp
220 in the strength of their current star formation, most are of low total
luminosity even while, apparently, at the peak of their star formation
rates.  This is in sharp contrast with an IRAS selected local sample of
galaxies (Kim and Sanders 1998) where a majority of those systems are
extremely luminous and massive.  The cluster starburst population appears
to be composed of a unique type of galaxy, possibly the progenitors of
present-day dwarf ellipticals and BCD's.  Deeper {\it Hubble Space
Telescope} (HST) observations by Koo \etal (1997), Oemler \etal (1997) and
Couch \etal (1998) have shown that many of the starbursts in distant
clusters tend to be small objects whose final state is likely to be just
such a dwarf galaxy.  Further support for this view comes from the
observations that dwarf galaxies in Virgo, Fornax and Coma have undergone
recent episodes of star formation in the last 4 to 5 Gyrs (Bothun and
Caldwell 1984), which corresponds to the epoch of the starburst population
in all three of the intermediate redshift cluster we have studied.

\section{CONCLUSIONS}

The trends with respect to cluster populations for A115 (southern
subcluster) and A2283 (combined with the data for A2218 and A2317) are
shown in Figure 9.  The data for A2317 are published in Rakos, Odell and
Schombert (1997) and the data for A2218 will be published in Rakos,
Dominis and Steindling (2000).  Four parameters are plotted with respect
to the elliptical/S0 fraction; the fraction of the blue population
($f_B$), the fraction of starburst galaxies, the richness of the cluster
($C$) and the radius of where the maximum value of $f_B$ occurs.

The fraction of the blue population drops inversely with the E/S0
percentage.  This is not too surprising since a majority of the red
population is classified photometrically as E/S0's.  However, the relation
is nearly 1-to-1, suggesting that it is conversion of the E/S0 objects
into the blue population that comprises the Butcher-Oemler effect.  Since
our previous work on present-day ellipticals and the color evolution of
ellipticals (Schombert \etal 1993, Rakos and Schombert 1995) demonstrates
that their colors have a narrow range and closely match the models for a
passively evolving population with a formation redshift of 5, we must
conclude that it is primarily the S0 galaxies that are the normal
starforming objects of the Butcher-Oemler population.  This scenario
matches well to our expectations of gas depletion, where large bulge S0's
exhaust their supply first, followed by Sa's, then Sb's in the near
future.  We can assume that a gas-rich, proto-S0 undergoing star formation
rates of around 1 $M_{\sun}$ yrs$^{-1}$ will produce spiral structure and
have morphologies similar to early-type spirals as revealed by HST
imaging.

The fraction of starbursts is only weakly related to the E/S0 fraction
confirming the split impact that the reddened starburst galaxies have on
the color distribution of cluster populations.  This is a new phenomenon
for distant clusters and, while the same mechanisms may be at work for the
dwarf starbursts, we believe this is a separate event from the
Butcher-Oemler effect.

The radius of the peak density of the blue population is also uncorrelated
with any population fraction or color fraction of the clusters.  This fact
is illuminating in the sense that distribution of the blue population in
the red clusters (A115 and A2218) is similar to that of the blue clusters
(A2283 and A2317).  If tidal forces are responsible for the blue
population, then there will be a mix of effects due to close encounters
with other galaxies and the mean cluster tidal field which will only be
weakly dependent on the cluster morphology.  If ram pressure stripping is
important, then this radius should be correlated with the x-ray profile of
the cluster.

One of the primary results of this study is that there appears to exist a
{\it duality} with respect to cluster population studies by morphology versus
photometric classification.  Morphological studies of intermediate
redshift clusters (Oemler, Dressler and Butcher 1997, Dressler \etal 1997)
find the fraction of ellipticals to be similar to present-day clusters
with a sharp drop in the number of S0's and a proportional increase in the
number of spirals and irregulars.  There is also an increase in the number
of disturbed galaxies with redshift (Couch \etal 1998), galaxies with
tidal signatures of past encounters.  On the other hand, photometric
classification (such as this study and Dressler \etal 1999) find a
population of bright blue galaxies with normal star formation rates (the
Butcher-Oemler effect) and a starburst population, dominated by low mass
galaxies. But, the characteristics of these populations differ in their
luminosity and geographical distributions compared to their morphological
counterparts.  This is particularly evident in magnitude diagrams such as
Figure 5 compared to similar diagrams for nearby, rich clusters.  {\it The
underlying consequence of this duality is the decoupling of the presence
of star formation of the properties of Hubble types with redshift} (i.e.
the existence of blue S0's, starbursting dwarfs, disturbed spirals at high
redshift).

How this decoupling takes place is due to a separation of formation
events (such as the density of the protogalaxy, the fraction of dark to
baryonic matter) and the later effects of environment.  Clearly, local density
plays an important role in the evolution of a galaxy.
As shown in Hashimoto \etal 1998, galaxies at intermediate densities have higher
star formation rates than galaxies located in high dense regions (such as
the core of a rich cluster).  If the dominate process were galaxy
encounters that induce star formation, then this correlation would 
be expected as intermediate density environments (such as loose groups) have the
appropriate mix of galaxy density and low velocities to maximize tidal
effects (Hashimoto \etal 1998).  However, a competing process, that a pure
density analysis does not take into account, is the possibly of ram
pressure effects from the hot intracluster gas identified with every rich
cluster of galaxies.  There is clear evidence of such stripping ongoing in
nearby clusters, such as Virgo, based on the HI and CO images of cluster
spirals (Kenney and Young 1989).  

We conclude that the Butcher-Oemler effect in rich clusters is the
phenomenon of starforming S0's being ram pressure stripped as they
encounter the hot gas in the core of the cluster.  The elimination of the
atomic gas prematurely ceases star formation (although the question
remains of the interplay with the molecular gas) and the S0's disks age to
the population we see today.  On the other hand, the dwarf galaxy interactions
in the outer regions of the cluster (where the velocities are lower)
produce the starburst population.  Both the blue and starburst population
avoid the core of the cluster, but each for different reasons, one being
intrinsic, the other environmental.

\acknowledgements
The authors wish to thank the directors and staffs of KPNO, Steward and
Wise Observatories for granting time for this project.  Financial support
from Austrian Fonds zur Foerderung der Wissenschaftlichen Forschung is
also gratefully acknowledged.  This research has made use of the NASA/IPAC
Extragalactic Database (NED) which is operated by the Jet Propulsion
Laboratory, California Institute of Technology, under contract with the
National Aeronautics and Space Administration.

\clearpage 

\figcaption{Two color diagrams and tracks for redshift effects.  The
shaded areas represent the regions occupied by ellipticals and spirals.
The two tracks are the observed colors for a typical elliptical spectrum
(NGC 3379) and a starforming spiral (NGC 6643) for redshifts ranging from
$-0.10$ to 0.10.  Background and foreground galaxies (beyond $-0.05$ and
0.05 of the cluster mean) can be identified by their deviant color
indices.  Additional membership information from the $vz-bz$ colors are not
shown.  }

\figcaption{Two color diagrams for template galaxies.  The
two color indices, $vz-yz$ and $bz-yz$ (roughly 4250\AA\ - 5500\AA\ and
4750\AA\ - 5500\AA) are shown for the four photometric classes based
on template data discussed in RMS96.  The solid line in each panel is the normal
starforming sequence.  The spiral templates are divided into Sa's
(crosses), Sb (solid diamonds) and Sc's (squares).  The starburst
templates are divided into normal (solid diamonds) and
interactions/mergers (crosses) based on disturbed morphology.  As
discussed in the text, the ellipticals and S0's form a distinct
grouping in multi-color space at the red end.  Spirals form a linear
sequence with the early-type spirals (large red bulge) overlapping the
E/S0 region and the late-type spirals overlapping the starburst
region.  Seyferts form a sequence above the spirals, indicating the
contribution from a non-thermal component to their optical
luminosity.  The starbursts lie below the spiral sequence depending
on the amount of reddening present.} 

\figcaption{Two color and $mz$ diagrams for A115.  The four
photometric classes are displayed in the $vz-yz$, $bz-yz$ and $mz$
planes.  The solid line in the $vz-yz$, $bz-yz$ diagram is the spiral
sequence from Figure 2.  The solid lines in the $vz-yz$, $mz$ diagram
demark the the classification boundaries from RMS96.  A typical error
bar for the faintest galaxies is shown.  Although a majority of the
galaxies are red, non-starforming systems, there are a significant
number of bright, blue spirals (the Butcher-Oemler effect) as well as
a low luminosity population of starburst galaxies.}

\figcaption{Two color and $mz$ diagrams for A2283.  The four
photometric classes are displayed in the $vz-yz$, $bz-yz$ and $mz$
planes.  Lines and symbols are the same as Figure 3.
A2283 is a high $f_B$ cluster.  The diagrams show that a majority of
the blue galaxies are normal starforming galaxies (spirals).  }

\figcaption{Luminosity distribution for the three primary population
types in A115, A2283 and A2317.  The composite luminosity histograms
in the $M_{5500}$ magnitude system (approximately Johnson $V$) are
shown for all three clusters.  The ellipticals/S0s display typical
luminosities for a rich cluster.  However, the spirals/irregulars are
strongly peaked at high luminosity indicating their dominant
contribution to the Butcher-Oemler effect.  The starburst galaxies
display an opposite behavior, rising to a peak at low luminosities.
Incompleteness is estimated to begin at $M_{5500} = -19$.}

\figcaption{The population fractions for the blue clusters (A2283 and
A2317) versus the red cluster (A115).  The three primary population
types (E/S0, spiral and starburst) are divided into 4 luminosity
bins.  The solid lines show the change in $f_B$ with luminosity.
Both blue and red clusters display similar traits with respect to the
drop then rise of $f_B$ and the rise of the fraction of starburst
galaxies with decreasing luminosity, despite their differences in
global population types.  The blue population is dominated by normal
starforming objects at high luminosity, with an increasing
contribution from low mass starbursts. }

\figcaption{The color-magnitude relation for the red populations in
A115, A2283 and A2317.  The various symbols display the metallicity
color index, $vz-yz$, versus absolute magnitude, $M_{5500}$.  The
solid line is a linear fit to the A115 and A2283 data.  The dotted
line is from the relationship for present-day ellipticals (Schombert
\etal 1993).  While the scatter is high, the differences in the slopes
of the three clusters is clear.  The large range for the brightest
red galaxies may be due to HB contributions to the integrated light,
or blue S0's evolving into normal S0 colors.}

\figcaption{The radial distribution of the blue and red populations
for A2283.  The blue and red populations are divided into four radial
bins measured from the geometric cluster center.  The blue population
is clearly deficient in the cluster core and peaks 0.5 Mpc from the
cluster center.  The red population dominates the cluster core by
number.}

\figcaption{Summary of cluster properties for A115, A2218, A2283 and
A2317.  Four parameters are plotted with respect with respect to the
E/S0 fraction for each cluster; the fraction of the blue population
($f_B$), the fraction of starburst galaxies, the richness of the
cluster ($C$) and the radius of where the maximum value of $f_B$
occurs.}

\clearpage
\begin{deluxetable}{lccccccccc}
\tablecolumns{10}
\small
\tablewidth{0pt}
\tablecaption{Clusters Properties}
\tablehead{\colhead{Cluster} & \colhead{RA (1950) Dec} & \colhead{BM type}
& \colhead{C} & \colhead{R} & \colhead{z} & \colhead{$f_B$} &
\colhead{E/S0\%} & \colhead{Starburst \%} & \colhead{max $f_B$ radius} }
\startdata
A115  & 0056.0+2619 & III &  174 & 3 &  0.197 & 0.13 & 67$\pm$8 & 17$\pm$4 & 0.45 \nl
A2218 & 1635.9+6612 & II  &  214 & 4 &  0.171 & 0.23 & 65$\pm$7 & 10$\pm$3 & 0.50 \nl
A2283 & 1744.5+6938 & I   &   65 & 1 &  0.183 & 0.30 & 49$\pm$8 & 22$\pm$5 & 0.55 \nl
A2317 & 1908.3+6903 & II  &  186 & 3 &  0.211 & 0.35 & 42$\pm$7 & 22$\pm$4 & 0.30 \nl
\enddata
\end{deluxetable}

\begin{deluxetable}{lccc}
\tablecolumns{4}
\small
\tablewidth{0pt}
\tablecaption{A115 Cluster Population}
\tablehead{\colhead{Type} & \colhead{Blue} & \colhead{Red} & \colhead{Total} }
\startdata
E/S0       &   1 ( 8\%) &  67 (78\%) &   68 (68$\pm$8\%) \nl
Sp/Irr     &   1 ( 8\%) &  10 (11\%) &   11 (11$\pm$3\%) \nl
Sy/AGN     &   4 (20\%) &   0 ( 0\%) &    4 ( 4$\pm$2\%) \nl
Starburst  &   7 (54\%) &  10 (11\%) &   17 (17$\pm$4\%) \nl
Total      &  13 (13$\pm$3\%) &  87 (87$\pm$9\%) &  100 \nl
\enddata
\end{deluxetable}

\begin{deluxetable}{lccc}
\tablecolumns{4}
\small
\tablewidth{0pt}
\tablecaption{A2283 Cluster Population}
\tablehead{\colhead{Type} & \colhead{Blue} & \colhead{Red} & \colhead{Total} }
\startdata
E/S0       &  1 ( 4\%) &  38 (69\%) &  39 (49$\pm$8\%) \nl
Sp/Irr     & 11 (46\%) &   9 (16\%) &  20 (25$\pm$6\%) \nl
Sy/AGN     &  3 (13\%) &   0 ( 0\%) &   3 ( 4$\pm$2\%) \nl
Starburst  &  9 (38\%) &   8 (15\%) &  17 (22$\pm$5\%) \nl
Total      & 24 (30$\pm$6\%) &  55 (70$\pm$9\%) &  79 \nl
\enddata
\end{deluxetable}

\pagestyle{empty}
\clearpage
\begin{figure}
\plotfiddle{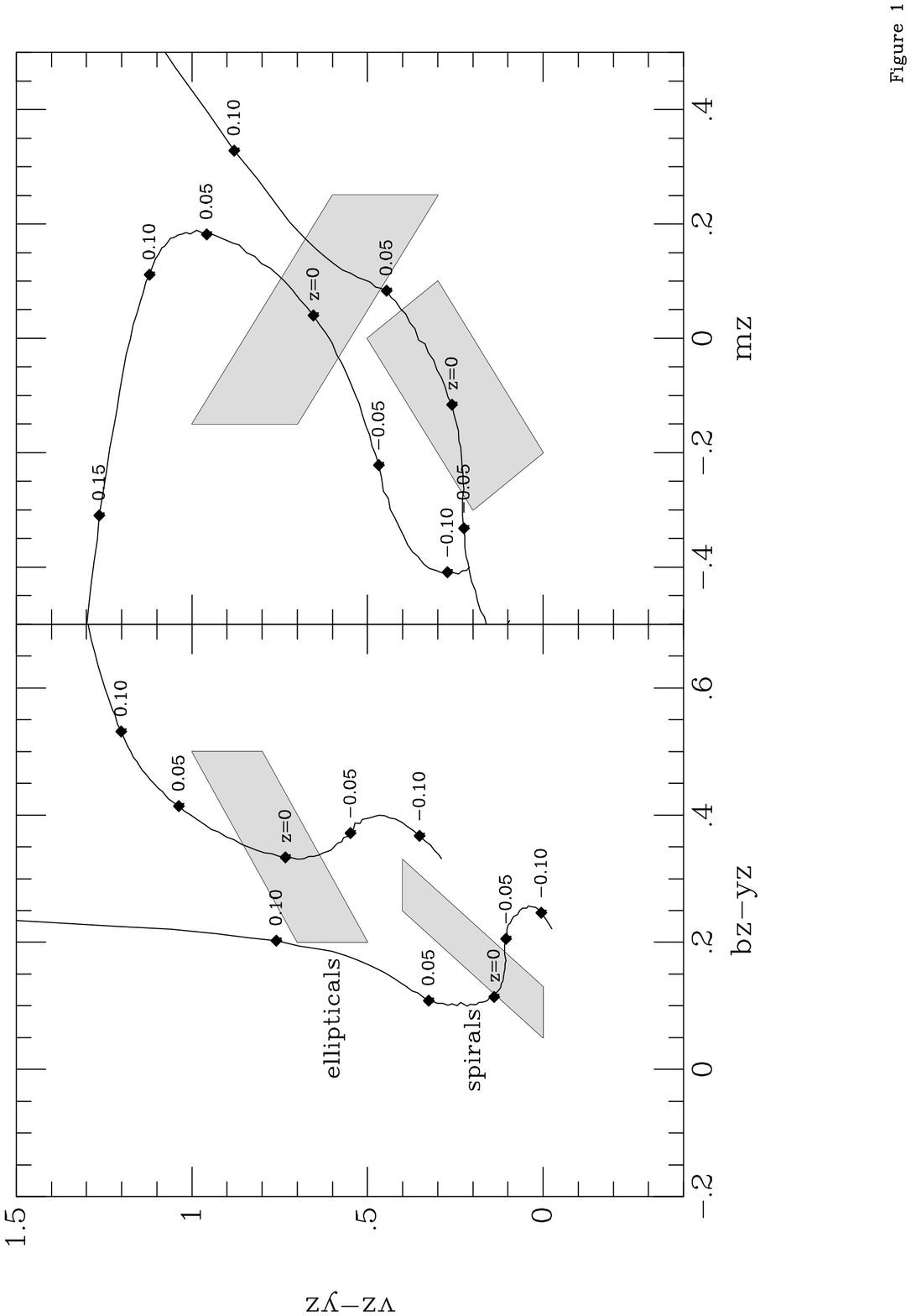}{11.5truein}{0}{100}{100}{-310}{170} \end{figure}

\clearpage
\begin{figure}
\plotfiddle{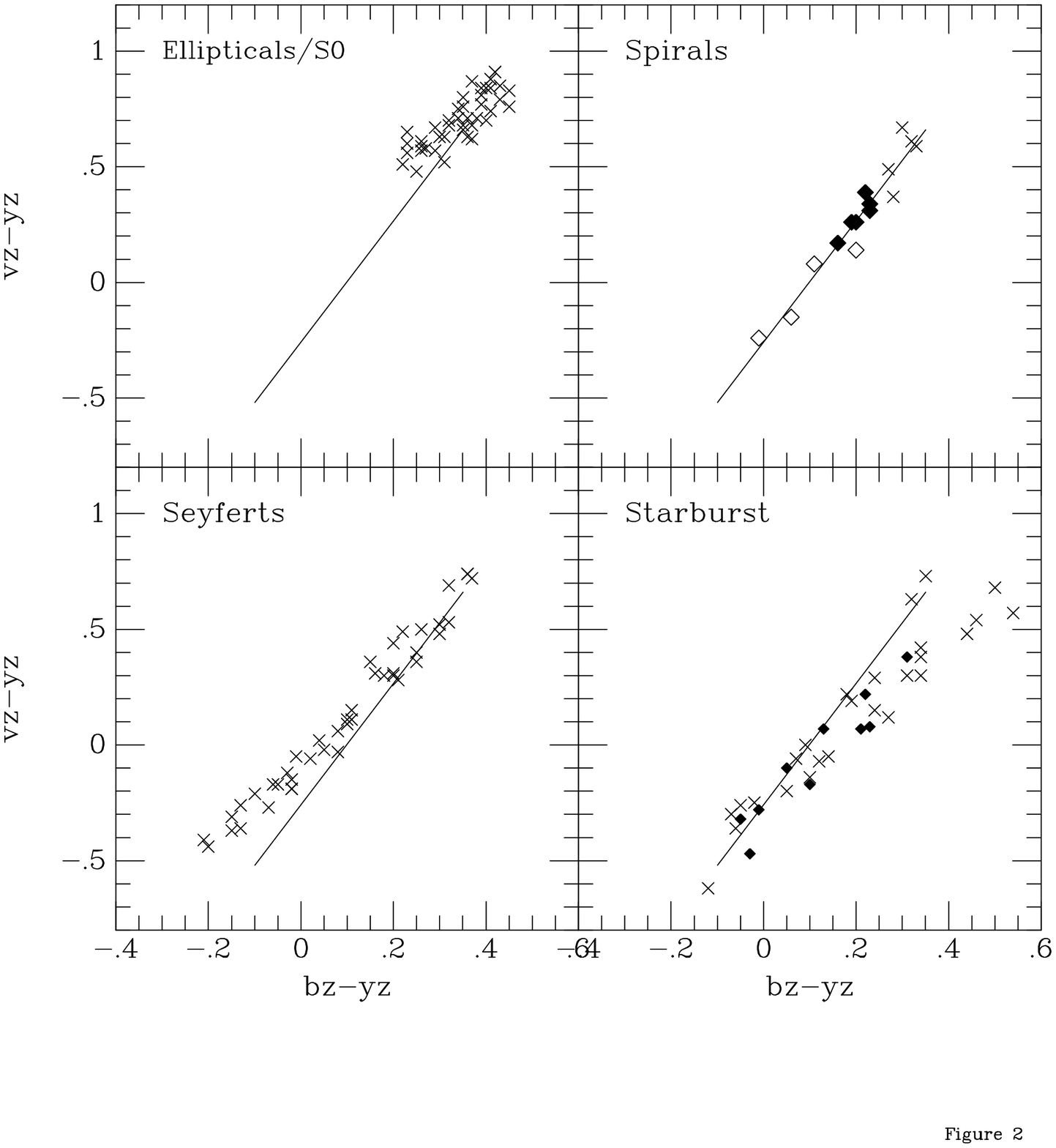}{11.5truein}{0}{100}{100}{-310}{170} \end{figure}

\clearpage
\begin{figure}
\plotfiddle{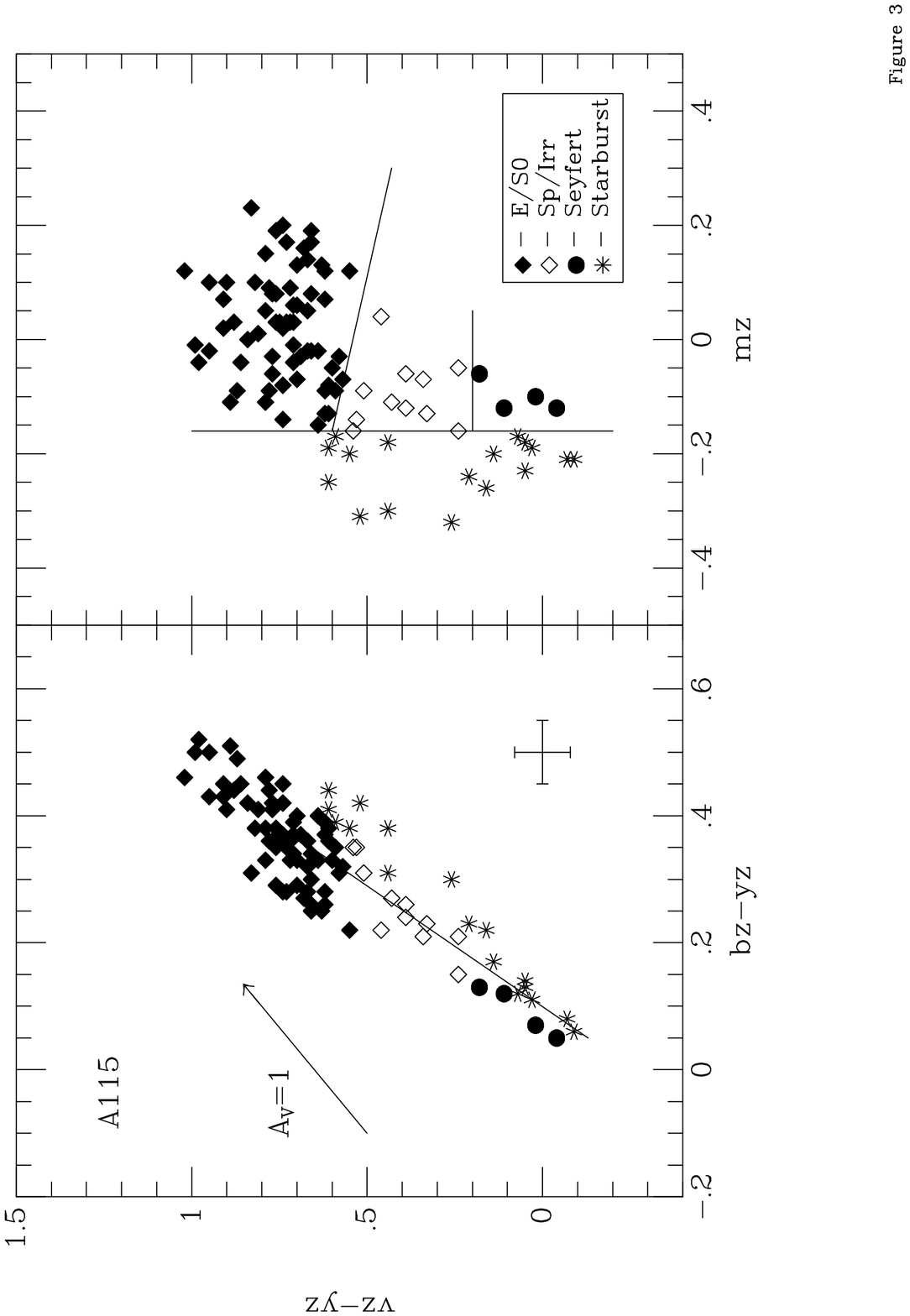}{11.5truein}{0}{100}{100}{-310}{170} \end{figure}

\clearpage
\begin{figure}
\plotfiddle{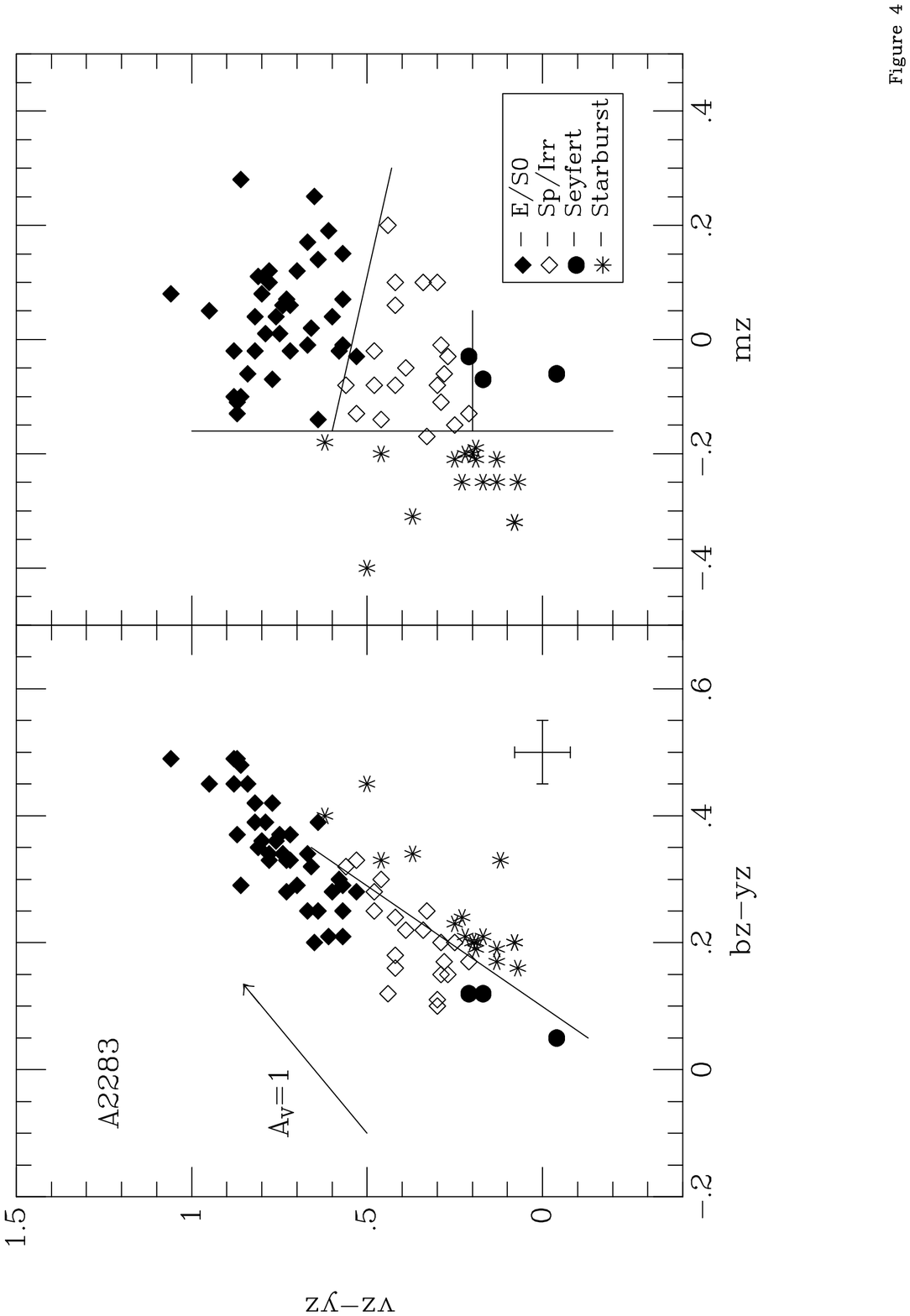}{11.5truein}{0}{100}{100}{-310}{170} \end{figure}

\clearpage
\begin{figure}
\plotfiddle{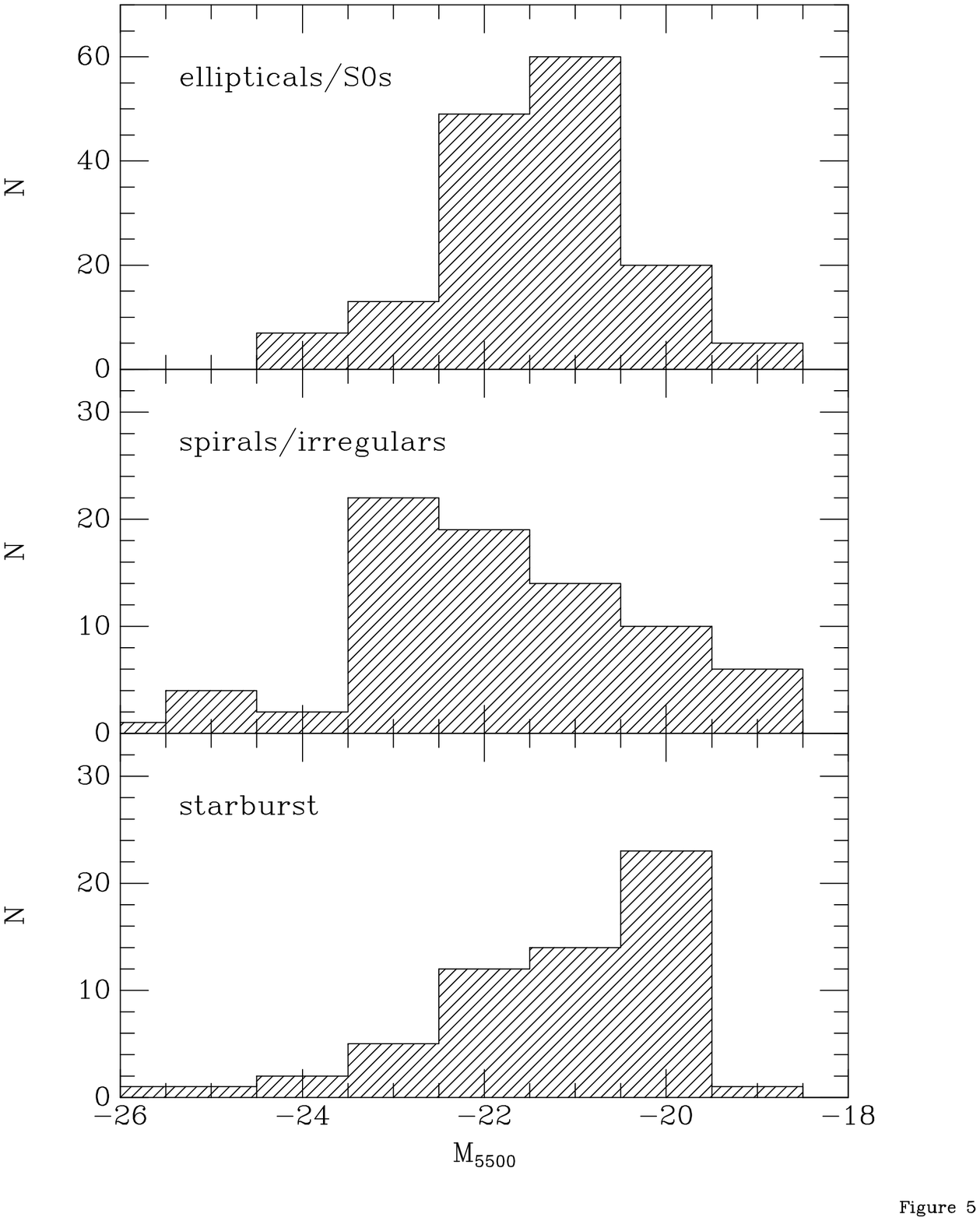}{11.5truein}{0}{100}{100}{-310}{170} \end{figure}

\clearpage
\begin{figure}
\plotfiddle{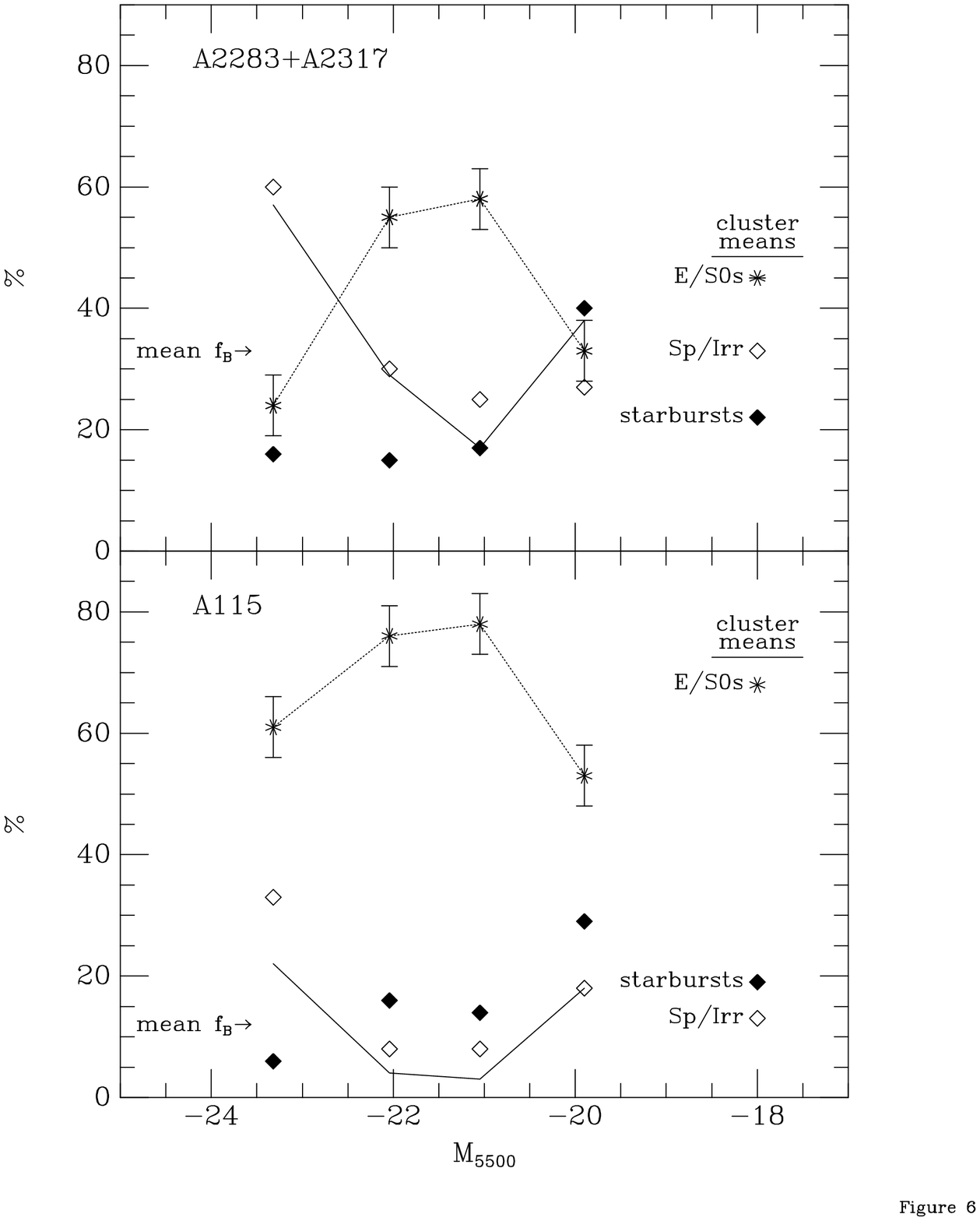}{11.5truein}{0}{100}{100}{-310}{170} \end{figure}

\clearpage
\begin{figure}
\plotfiddle{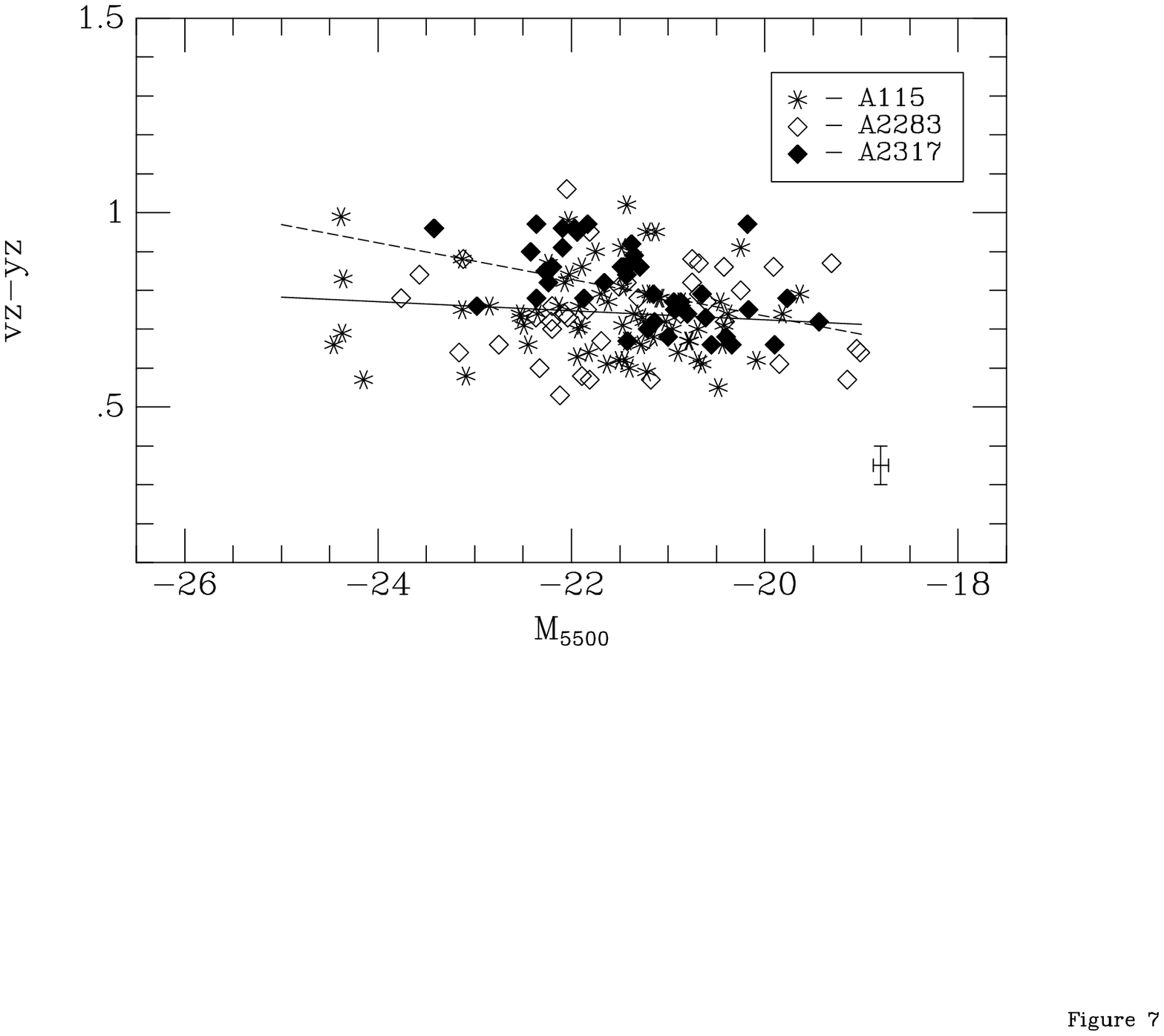}{11.5truein}{0}{100}{100}{-310}{170} \end{figure}

\clearpage
\begin{figure}
\plotfiddle{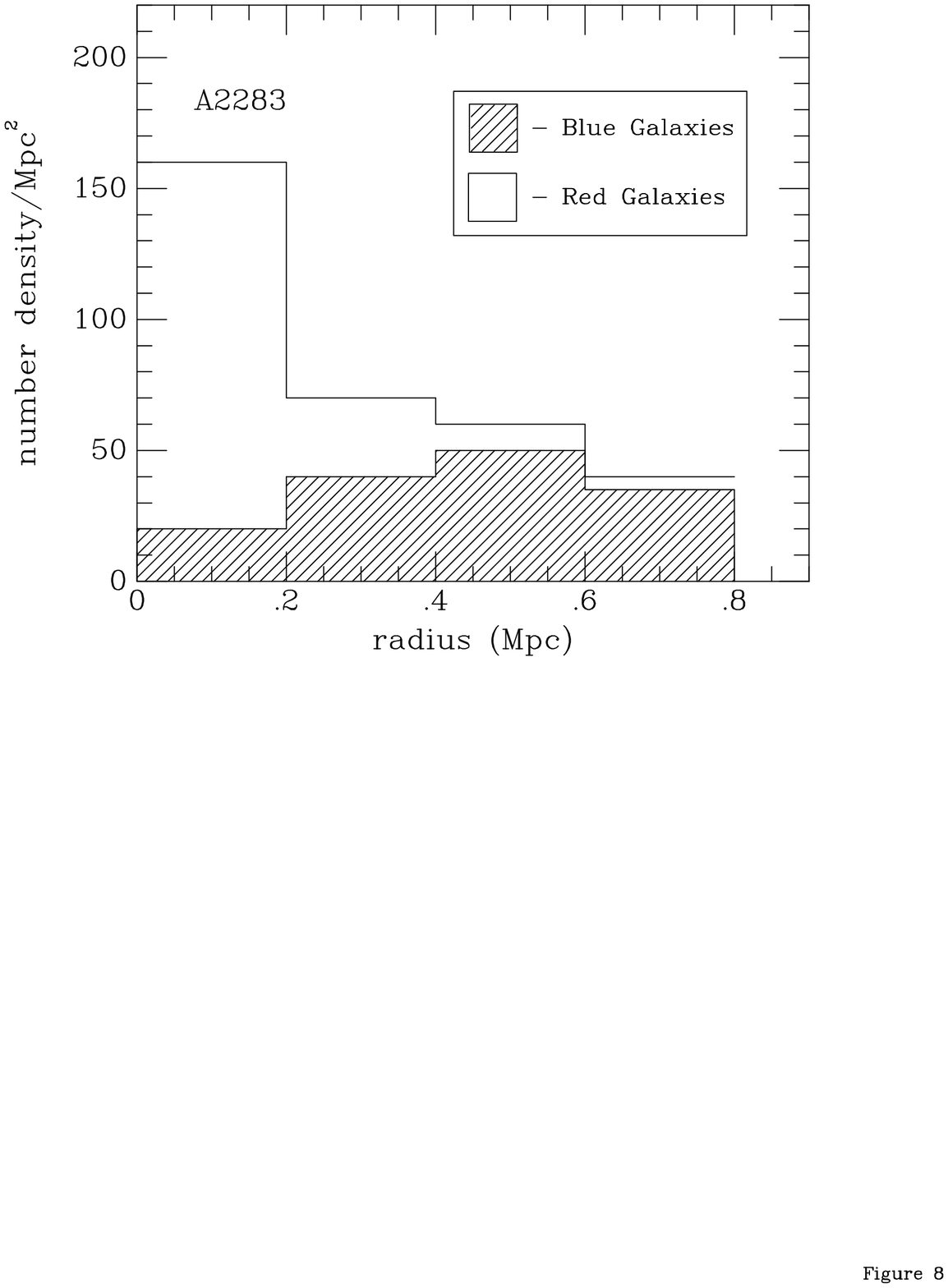}{11.5truein}{0}{100}{100}{-310}{170} \end{figure}

\clearpage
\begin{figure}
\plotfiddle{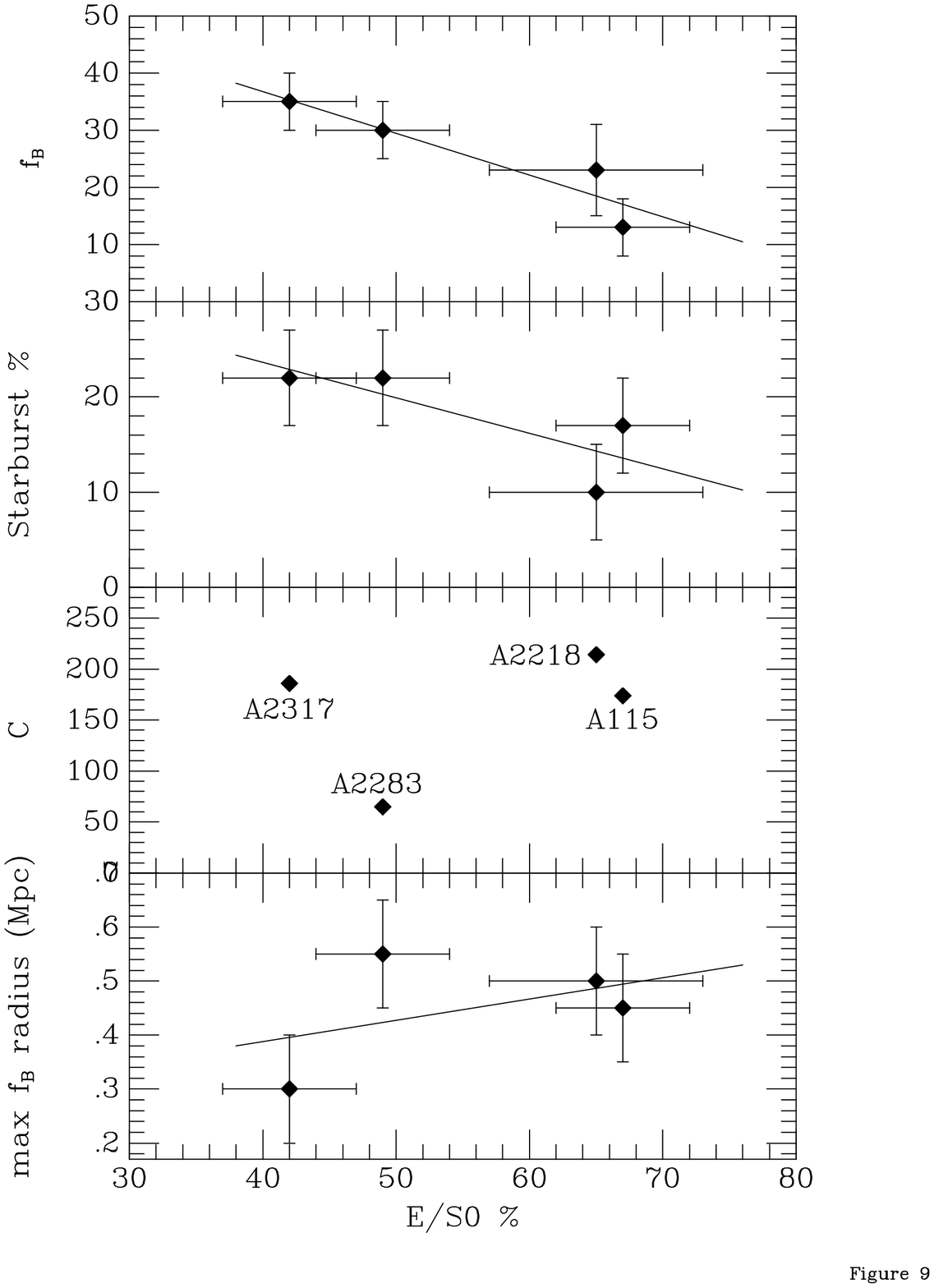}{11.5truein}{0}{100}{100}{-310}{170} \end{figure}

\end{document}